\documentclass[conference]{IEEEtran}
\usepackage{amsmath}
\usepackage{epsfig}
\usepackage{color}
\usepackage[normalem]{ulem}
\usepackage{amssymb}
\usepackage{latexsym}
\usepackage{graphicx}
\usepackage{cite}

\title{Analytic Performance Evaluation of Underlay Relay Cognitive Networks with Channel Estimation Errors}

\author{
Khuong~Ho-Van$^{1}$,~Paschalis~C.~Sofotasios$^{2}$,~Son~Vo
Que$^{1}$,~Tuan~Dang~Anh$^{1}$,~Thai~Pham~Quang$^{1}$,~Lien~Pham
Hong$^{3}$
\\\\
\begin{normalsize}
$^{1}$Department of Telecommunications Engineering, HoChiMinh City
University of Technology, HoChiMinh City, Vietnam.
\end{normalsize}
\\
\begin{normalsize}
e-mail: khuong.hovan@yahoo.ca, \{sonvq, datuang\}@hcmut.edu.vn,
pqthai.hcmut@gmail.com
\end{normalsize}
\\
\begin{normalsize}
$^{2}$School of Electronic and Electrical Engineering, University of
Leeds, LS$2$ $9$JT, Leeds, United Kingdom.
\end{normalsize}
\\
\begin{normalsize}
e-mail: p.sofotasios@leeds.ac.uk
\end{normalsize}
\\
\begin{normalsize}
$^{3}$Department of Electrical and Electronics Engineering,
University of Technical Education, HoChiMinh City, Vietnam.
\end{normalsize}
\\
\begin{normalsize}
e-mail: phamhonglien2005@gmail.com
\end{normalsize}
 }

\begin{document}

\maketitle

\begin{abstract}
This paper evaluates the bit error rate (BER) performance of underlay relay cognitive networks with decode-and-forward (DF)
relays in arbitrary number of hops over Rayleigh fading  with channel estimation errors. 
In order to facilitate the performance evaluation analytically
 we derive a novel exact closed-form representation for the corresponding BER which is validated through
 extensive comparisons with results from Monte-Carlo simulations. The proposed expression involved well known elementary and special functions which render its computational realization rather simple and straightforward. As a result, the need for laborious, energy exhaustive and time-consuming computer simulations can be ultimately omitted.
Numerous results illustrate that the performance of underlay relay
cognitive networks is, as expected, significantly degraded by channel estimation
errors and that is highly dependent upon of both the network topology and the number of
hops.
\end{abstract}


\maketitle

\section{Introduction}
\label{intro}

It was recently pointed out by a spectrum usage survey from the
Federal Communications Commission (FCC), that the current licensed
spectrum situation is significantly under-utilized \cite{FCC2002}.
Contrary to that, the current availability of spectrum resources for
most emerging wireless applications such as video calling, online
high-definition video streaming, high-speed Internet access through
mobile devices, etc. are particularly scarce. In an attempt to
improve the spectrum utilization in wireless communication systems,
cognitive radio (CR) technology was proposed as a promising
technology \cite{Mitola2000, Paschalis_1, Paschalis_2, Paschalis_3, Theo, Paschalis_4, Paschalis_5}. In cognitive radio, secondary
users-SUs (or unlicensed users) are generally allowed to use the
licensed band primarily allotted to primary users-PUs (or licensed
users), unless their operation interferes with the established
communication of PUs. This operation can be realized in three
distinctive modes: underlay, overlay and interweave \cite{Lee2011}.
In the underlay mode, SUs are allowed to use the spectrum when the
interference caused by SUs on PUs is within a tolerated range by
PUs. This mode is more preferable than its two counterparts thanks to its low
implementation complexity \cite{Goldsmith2009}.

Due to the interference power constraint imposed on SUs operating in
the underlay mode, their transmit power is limited and as such,
their transmission range is reduced substantially. To overcome this
constraint, SUs can apply relaying techniques, which take advantage of
shorter range communication that results to lower path loss effects. Among various
relaying techniques, decode-and-forward (DF) and amplify-and-forward
(AF) deployments have been extensively investigated \cite{Laneman2004}. In DF,
each relay decodes information from the source, re-encodes it, and
forwards it to the destination. In AF, each relay simply
amplifies the received signal and forwards it to the destination.
Due to its capability of regenerating noise-free relayed signals, DF
is employed in this paper.

This paper investigates underlay DF multi-hop cognitive networks
with arbitrary number of hops. Most relevant works considering such
network deployments focus in outage probability analysis
\cite{Lee2011}, \cite{Jun, Caijun, Yan, khuong, Liping, Yang}, and
BER analysis\footnote{ The work in \cite{KhuongIEICE} derives an
approximate closed-form BER expression.} \cite{Hussain, Do,
KhuongIEICE} assuming perfect channel estimation and two-hop
communication. It is also recalled here that channel state
information (CSI) is essential for coherent detection; nevertheless,
existing channel estimators are unable to provide and guarantee
perfect CSI. As a consequence, the impact of imperfect CSI on the
system performance should be considered realistically.

In \cite{Sura2010}, the BER analysis for \emph{single-hop} cognitive
networks is presented under the assumption of imperfect CSI only
for SU-PU links. In \cite{Chen}, an exact outage probability
expression was proposed for \emph{AF dual-hop} cognitive networks. However, to the best of
our knowledge, the exact BER analysis for underlay \emph{DF $N$-hop}
cognitive networks, with $N$ being arbitrary integer, and
imperfect CSI on all wireless channels, has not been addressed in the open technical literature. Motivated by this, this paper is devoted to an analytic investigation of this topic by deriving a corresponding exact closed-form BER
expression. The derived expression is validated by extensive computer simulations and
is utilized in evaluating the corresponding system performance.

The structure of this paper is as follows: The next section presents
the system model and the CSI imperfection model. The BER analysis is
discussed in Section \ref{sec:PerAna} while simulated and analytical
results are presented in Section \ref{IllustrativeResults} for
derivation validity and performance evaluation. Finally, the paper
is concluded in Section~\ref{Conclusion}.

\section{System Model}
\label{sec:SystemModel}

\begin{figure}[!h]
\centering
\includegraphics[height=6cm, width=9cm]{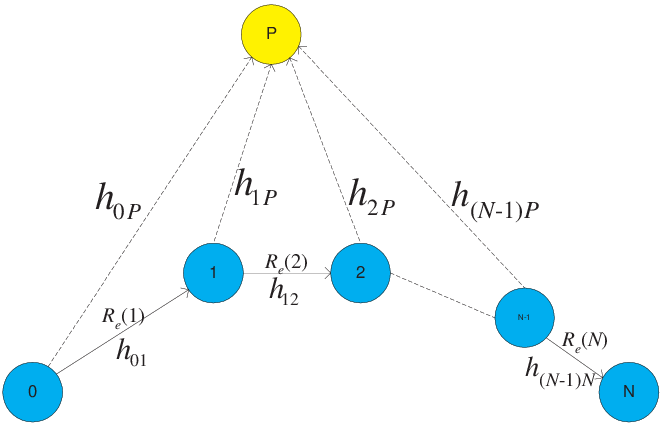}
\caption{System model.} \label{Fig_SM}
\end{figure}

The underlay cognitive DF multi-hop network model under
consideration is depicted in Fig.~\ref{Fig_SM}, where $N-1$
secondary relays (SRs) numbered from 1 to $N-1$ assist the
transmission of the secondary source (SS) 0 to the secondary
destination (SD) $N$. The SS and SRs use the same spectrum as a
primary user P. The direct communication between SS and SD is
bypassed, which is considered reasonable in scenarios where SS and
SD are too far apart or their communication link is blocked due to
severe shadowing and fading. We assume that the channel between any
pair of transmitter and receiver experiences independent block
frequency-flat Rayleigh fading i.e., frequency-flat fading is
invariant during one phase but independently changed from one to
another. Therefore, the channel coefficient between the transmitter
$t\in \{0, 1, \ldots, N-1\}$ and the receiver $r\in \{1, 2, \ldots,
N, P\}$ is $h_{tr} \sim \mathcal{CN}\left(0, {\eta
_{tr}}=d_{tr}^{-\alpha}\right)$ \footnote{$h \sim \mathcal{CN}(m,v)$
denotes an $m$-mean circular symmetric complex Gaussian random
variable with variance $v$.}, where $d_{tr}$ is the distance between
the two terminals and $\alpha$ is the path-loss exponent
\cite{Ahmed}.

An $N$-hop communication time interval consists of $N$ phases. In the
first phase, SS 0 transmits a sequence of $K$ modulated symbols
$\textbf{x}_0 = [x_0(1), x_0(2), ..., x_0(K)]$ with the symbol
energy, $B_0$ i.e., $E\{{\left| x_0(k) \right|^2}\}=B_0$ where
$E\{\cdot\}$ denotes the expectation and $k$ is the time index. SR
1 demodulates the received signal from SS 0 and re-modulates the
demodulated symbol as $\textbf{x}_1 = [x_1(1), x_1(2), ..., x_1(K)]$
with the symbol energy, $B_1$, before forwarding to SR 2 in the
second phase. The process continues until the signal reaches SD $N$.
Without the notation confusion, the time index is omitted in the
sequel and hence, the received signal through the hop $r$ can be
expressed as
\begin{eqnarray}
\label{y_r} {y_{tr}} = {h_{tr}}{x_t} + {n_{tr}},
\end{eqnarray}
where $y_{tr}$ denotes a signal received at the node $r$ from the
node $t=r-1$ and $n_{tr} \sim \mathcal{CN}(0, N_0)$ is additive
white Gaussian noise at the node $r$.

In the underlay relay cognitive networks (e.g., \cite{Liping},
\cite{Guo}), the SU $t$'s transmit power is limited such that the
interference imposed on PU is under control. Without CSI errors,
this interference constraint can be addressed as $B_t \leq
I_T/|h_{tP}|^2$ where $I_T$ is the maximum interference level that
PU still operates reliably. For the maximum transmission range, $B_t
= I_T/|h_{tP}|^2$ is set.
Following \cite{Amin, Seungyoup, Yi, Patel}, we choose the CSI
imperfection model as
\begin{eqnarray}
\label{h_tr} {h_{tr}} = {\widehat h_{tr}} + {\varepsilon_{tr}},
\end{eqnarray}
where $\widehat h_{tr}$ is the estimate of the $t-r$ channel and
$\varepsilon_{tr}$ is the CSI error.

We assume that $h_{tr}$ and $\widehat h_{tr}$ are jointly ergodic
and stationary Gaussian processes. Therefore, $\varepsilon_{tr} \sim
\mathcal{CN}\left(0, {\sigma _{tr}} \right)$ and $\widehat h_{tr}
\sim \mathcal{CN}\left(0, \frac{1}{\lambda_{tr}} = {\eta
_{tr}}-\sigma _{tr} \right)$ with $\sigma _{tr}$ representing the
quality of the channel estimator. For example \cite{Amin}, for the
linear-minimum-mean-square-error (LMMSE) estimator, $\sigma _{tr} =
E\left\{ {{{\left| {{h_{tr}}} \right|}^2}} \right\} - E\left\{
{{{\left| {{{\widehat h}_{tr}}} \right|}^2}} \right\} = 1/\left(
{{L_p}{{\bar \gamma }_{tr,training}} + 1} \right)$ where $L_p$ is
the number of pilot symbols, ${{\bar \gamma }_{tr,training}} =
E\left\{ {{\gamma _{tr,training}}} \right\} = {B_{t,training}}{\eta
_{tr}}/{N_0}$ is the average SNR of pilot symbols for the $t-r$
channel, and $B_{t,training}$ is the pilot power.

\section{Error Probability Analysis}
\label{sec:PerAna}

Due to CSI errors, the transmit power of the node $t$ is modified as
$B'_t = I_T/|\widehat h_{tP}|^2$. Then, there are two possibilities:
$|\widehat h_{tP}|^2 \leq |h_{tP}|^2$ and $|\widehat h_{tP}|^2 >
|h_{tP}|^2$. Setting the transmit power as $B'_t = I_T/|\widehat
h_{tP}|^2$ meets the interference power constraint for $|h_{tP}|^2
\leq |\widehat h_{tP}|^2$, since this case results in the
interference power as $B'_t|h_{tP}|^2=I_T|h_{tP}|^2/|\widehat
h_{tP}|^2\leq I_T$, but not for $|h_{tP}|^2
> |\widehat h_{tP}|^2$, since this case results in the interference
power as $B'_t|h_{tP}|^2=I_T|h_{tP}|^2/|\widehat h_{tP}|^2 > I_T$.
Given that $E\left\{ {{{\left| {{{\widehat h}_{tP}}} \right|}^2}}
\right\} \le E\left\{ {{{\left| {{h_{tP}}} \right|}^2}} \right\}$
where the equality holds for no CSI errors, on average such transmit
power setting may not meet the interference power constraint i.e.,
the interference at P is greater than $I_T$. Therefore, the primary
system performance may be severely degraded if the channel estimator
is not efficient. Consequently, in order to propose solutions to
interference reduction on primary systems, statistics of
interference at the PU receiver should be analyzed. The most
important statistics is the probability that the interference
exceeds $I_T$, namely the interference probability $P_I$ as used in
\cite{Chen}. It is noted that $P_I$ is derived for underlay
\emph{AF} \emph{dual-hop} cognitive networks \cite{Chen} and for
underlay \emph{single-hop} cognitive networks \cite{Sura2010} with
the CSI imperfection model slightly different\footnote{The CSI
imperfection model in \cite{Sura2010} and \cite{Chen} is ${\hat
h_{tr}} = {\rho _{tr}}{h_{tr}} + \sqrt {1 - \rho _{tr}^2}
{\varepsilon _{tr}}$ where $\rho_{tr}$ is the correlation
coefficient between $\hat h_{tr}$ and $h_{tr}$.}. Due to the space
limitation, the interference probability analysis is deferred to the
journal version of this paper. Instead, we focus on the BER analysis
for underlay relay cognitive networks. To this effect, using the CSI
imperfection model in (\ref{h_tr}), we rewrite (\ref{y_r}) as,
\begin{eqnarray}
\label{y_tr1} {y_{tr}} = \underbrace {{{\hat
h}_{tr}}{x_t}}_{{\rm{desired \
 signal}}} + \underbrace {{\varepsilon_{tr}}{x_t} + {n_{tr}}}_{{\rm{effective\
noise}}}.
\end{eqnarray}
According to (\ref{y_tr1}), the effective SNR of the $t-r$ channel
taking CSI errors into account is expressed as,
\begin{eqnarray}
\label{gamma _{tr,eff}} {\gamma _{tr}} &=& \frac{{{{\left|
{{{\widehat h}_{tr}}} \right|}^2}E\left\{ {{{\left| {{x_t}}
\right|}^2}} \right\}}}{{E\left\{ {{{\left| {{\varepsilon_{tr}}{x_t}
+ {n_{tr}}}
\right|}^2}} \right\}}} \nonumber\\
&=& \frac{{{B'_t}{{\left| {{{\widehat h}_{tr}}}
\right|}^2}}}{{{B'_t}{\sigma_{tr}} + {N_0}}} \nonumber\\
&=& \frac{{{{\left| {{{\widehat h}_{tr}}}
\right|}^2}}}{{{\sigma_{tr}}
+ {{\left| {{{\widehat h}_{tP}}} \right|}^2}/\mu }}\nonumber\\
&=&\frac{{{z_{tr}}}}{{{d_{tr}}}},
\end{eqnarray}
where ${z_{tr}} = {\left| {{{\widehat h}_{tr}}} \right|^2}$,
${d_{tr}} = {\sigma_{tr}} + {\left| {{{\widehat h}_{tP}}}
\right|^2}/\mu $, and $\mu = I_T/N_0$.

The average BER at the node $r$ for square $M$-QAM with $M = 2^q$
($q$ even) and rectangular $M$-QAM with $M = 2^q$ ($q$ odd)
modulation schemes\footnote{The average BER of other modulation
schemes such as $M$-PSK can be derived in the same approach.} is
expressed in \eqref{eq:BER_e2e_def} which is cited from \cite[eq.
(16)]{Cho2002} and \cite[eq. (22)]{Cho2002}, correspondingly. In
\eqref{eq:BER_e2e_def}, we define

\begin{figure*}[t!]
\begin{eqnarray}
\label{eq:BER_e2e_def} {R_e}\left( r \right) = \left\{
{\begin{array}{*{20}{c}}
{\int\limits_0^\infty  {\left\{ {\psi \left( {I,u,M;\gamma } \right) + \psi \left( {J,u,M;\gamma } \right)} \right\}{f_{{\gamma _{tr}}}}\left( \gamma  \right)d\gamma } }&{,{\rm{ }}q\ {\rm{ odd}}}\\
{2\int\limits_0^\infty  {\psi \left( {\sqrt M ,g,M;\gamma }
\right){f_{{\gamma _{tr}}}}\left( \gamma  \right)d\gamma } }&{,{\rm{
}}q\ {\rm{ even}}}
\end{array}} \right..
\end{eqnarray}
\end{figure*}
\begin{eqnarray}
g &=& \frac{3}{(M-1)},\\
u &=& \frac{6}{(I^2+J^2-2)},\\
I &=& 2^{(q-1)/2},\\
J &=& 2^{(q+1)/2},
\end{eqnarray}
and $\psi \left( {s,v,M;\gamma } \right)$ in \eqref{eq:xi_def} in
which $Q(.)$ is the Q-function \cite[eq. (1)]{C:Sofotasios_1},
\cite[eq. (10)]{C:Sofotasios_2}.

\begin{figure*}[t!]
\begin{eqnarray}
\label{eq:xi_def} \psi \left( {s,v,M;\gamma } \right) \buildrel
\Delta \over = \frac{2}{{s{{\log }_2}M}}\sum\limits_{k = 1}^{{{\log
}_2}s} {\sum\nolimits_{i = 0}^{\left( {1 - {2^{ - k}}} \right)s - 1}
{\frac{{{{\left( { - 1} \right)}^{\left\lfloor {\frac{{i{2^{k -
1}}}}{s}} \right\rfloor }}Q\left( {\sqrt {{{\left( {2i + 1}
\right)}^2}v\gamma } } \right)}}{{{{\left( {{2^{k - 1}} -
\left\lfloor {\frac{{i{2^{k - 1}}}}{s} + \frac{1}{2}} \right\rfloor
} \right)}^{ - 1}}}}} }.
\end{eqnarray}
\end{figure*}

Next, we derive $f_{\gamma_{tr}}(\gamma)$ in order to enable the derivation of an explicit
expression for (\ref{eq:BER_e2e_def}). Since $\widehat h_{tr} \sim
\mathcal{CN}\left(0, \frac{1}{\lambda_{tr}} \right)$ and $\widehat
h_{tP} \sim \mathcal{CN}\left(0, \frac{1}{\lambda_{tP}} \right)$,
the probability density functions (pdf's) of ${{z_{tr}}}$ and
${{d_{tr}}}$ are ${f_{{z_{tr}}}}\left( x \right) = {\lambda
_{tr}}{e^{ - {\lambda _{tr}}x}}$ and ${f_{{d_{tr}}}}\left( x \right)
= {\lambda _{tP}}\mu {e^{ - {\lambda _{tP}}\mu \left( {x -
{\sigma_{tr}}} \right)}}$, respectively. As a result, the pdf of
${\gamma _{tr}} = {z_{tr}}/{d_{tr}}$ in (\ref{gamma _{tr,eff}}) is
given as \cite[eq. (6-60)]{Athanasios}
\begin{eqnarray}
\label{eq:f_gamma}
{f_{{\gamma _{tr}}}}\left( x \right) &=& \int\limits_0^\infty  y {f_{{z_{tr}}}}\left( {yx} \right){f_{{d_{tr}}}}\left( y \right)dy\nonumber\\
 &=& \frac{{{\kappa _{tr}}\mu {e^{{\lambda _{tP}}\mu {\sigma_{tr}}}}}}{{{{\left( {x + {\kappa _{tr}}\mu }
 \right)}^2}}},
\end{eqnarray}
where ${\kappa _{tr}} = {\lambda _{tP}}/{\lambda _{tr}}$.

Inserting (\ref{eq:f_gamma}) into (\ref{eq:BER_e2e_def}) yields,
\begin{eqnarray}
\label{Eq:Re_def} {R_e}\left( r \right) = \left\{
{\begin{array}{*{20}{c}}
{\theta \left( {I,u,{W_{tr}}} \right) + \theta \left( {J,u,{W_{tr}}} \right)}&{,{\rm{ }}q\ {\rm{ odd}}}\\
{2\theta \left( {\sqrt M ,g,{W_{tr}}} \right)}&{,{\rm{ }}q\ {\rm{
even}}}
\end{array}} \right.
\end{eqnarray}
where ${W_{tr}} = \left\{ M, \kappa_{tr}, \mu, \lambda_{tP}, \sigma
_{tr}\right\}$ is a set of parameters and $\theta \left(
{s,v,{W_{tr}}} \right)$ is defined in \eqref{eq:theta_def}. Also,
$\zeta\left( {\beta ,a} \right)$ in \eqref{eq:theta_def} is defined
as

\begin{figure*}[t!]
\begin{eqnarray}
\label{eq:theta_def} \theta \left( {s,v,{W_{tr}}} \right) \buildrel
\Delta \over = \frac{2}{{s{{\log }_2}M}}\sum\limits_{k = 1}^{{{\log
}_2}s} {\sum\nolimits_{i = 0}^{\left( {1 - {2^{ - k}}} \right)s - 1}
{\frac{{{{\left( { - 1} \right)}^{\left\lfloor {\frac{{i{2^{k -
1}}}}{s}} \right\rfloor }}{\kappa _{tr}}\mu {e^{{\lambda _{tP}}\mu
{\sigma _{tr}}}}\zeta \left( {{{\left( {2i + 1} \right)}^2}v,{\kappa
_{tP}}\mu } \right)}}{{{{\left( {{2^{k - 1}} - \left\lfloor
{\frac{{i{2^{k - 1}}}}{s} + \frac{1}{2}} \right\rfloor } \right)}^{
- 1}}}}} }.
\end{eqnarray}
\end{figure*}
\begin{eqnarray}
\label{Idef} \zeta\left( {\beta ,a} \right) = \int\limits_0^\infty
{\frac{{Q\left( {\sqrt {\beta x} } \right)}}{{{{\left( {x + a}
\right)}^2}}}} dx.
\end{eqnarray}

Applying the integration by parts, we obtain the closed-form of
$\zeta\left( {\beta ,a} \right)$ as follows,
\begin{eqnarray}
\label{Iresult}
\zeta\left( {\beta ,a} \right) &=& \frac{1}{{2a}} - \frac{{\sqrt \beta  }}{{2\sqrt {2\pi } }}\int\limits_0^\infty  {\frac{{{e^{ - \frac{{\beta x}}{2}}}}}{{\left( {x + a} \right)\sqrt x }}} dx\nonumber\\
 &=& \frac{1}{{2a}} - \frac{{\sqrt \beta  {e^{\frac{{\beta a}}{2}}}}}{{2\sqrt {2\pi } }}\int\limits_a^\infty  {\frac{{{e^{ - \frac{{\beta y}}{2}}}}}{{y\sqrt {y - a} }}} dy\nonumber\\
 &=& \frac{1}{{2a}} - \sqrt {\frac{{\beta \pi }}{{2a}}} \frac{{{e^{\frac{{\beta a}}{2}}}}}{2}\left[ {1 - erf\left( {\sqrt {\frac{{\beta a}}{2}} } \right)} \right],\nonumber\\
\end{eqnarray}
where $erf\left( x \right) = \frac{2}{{\sqrt \pi }}\int\limits_0^x
{{e^{ - {t^2}}}dt} $ is the error function \cite[eq.
(8.250.1)]{Gradshteyn} and the closed-form expression of the
integral in the second equality is deduced with the aid of \cite[eq.
(3.363.2)]{Gradshteyn}.

Given the set of the average BERs of all hops $\{R_e(1),\cdots,
R_e(N)\}$, the exact closed-form average BER of the underlay DF
multi-hop cognitive networks is expressed as \cite[eq.
(9)]{Morgado2011}
\begin{eqnarray}
{R_e} = \mathop \sum \limits_{n = 1}^N \left[ {{R_e}\left( n
\right)\mathop \prod \limits_{j = n + 1}^N \left( {1 - 2{R_e}\left(
j \right)} \right)} \right].
\end{eqnarray}

\section{Numerical Results}
\label{IllustrativeResults}

\begin{figure}
\centering
\includegraphics{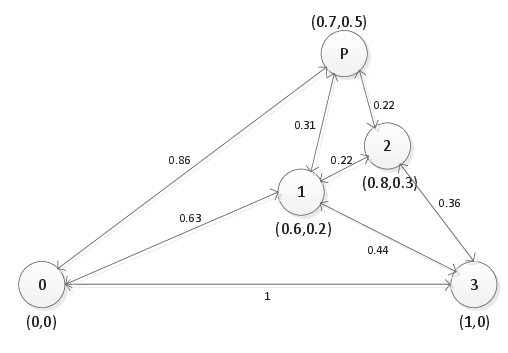}
\caption{Network topology.} \label{Fig:Sim}
\end{figure}

\begin{figure}
\centering
\includegraphics[height=9.5cm, width=9.5cm]{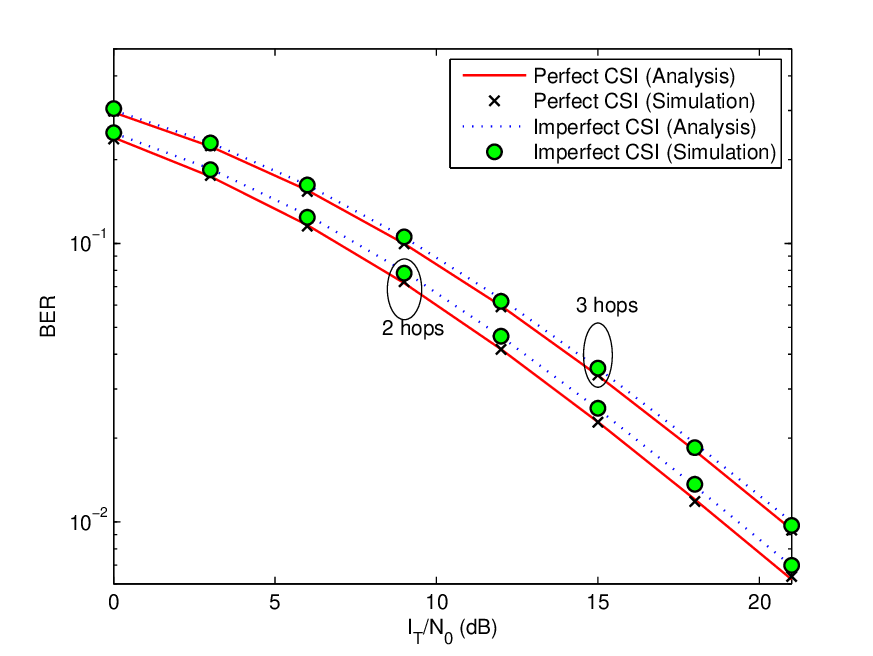}
\caption{BER versus $I_T/N_0$ (2-QAM).} \label{Figure12QAM}
\end{figure}

\begin{figure}
\centering
\includegraphics[height=9.5cm, width=9.5cm]{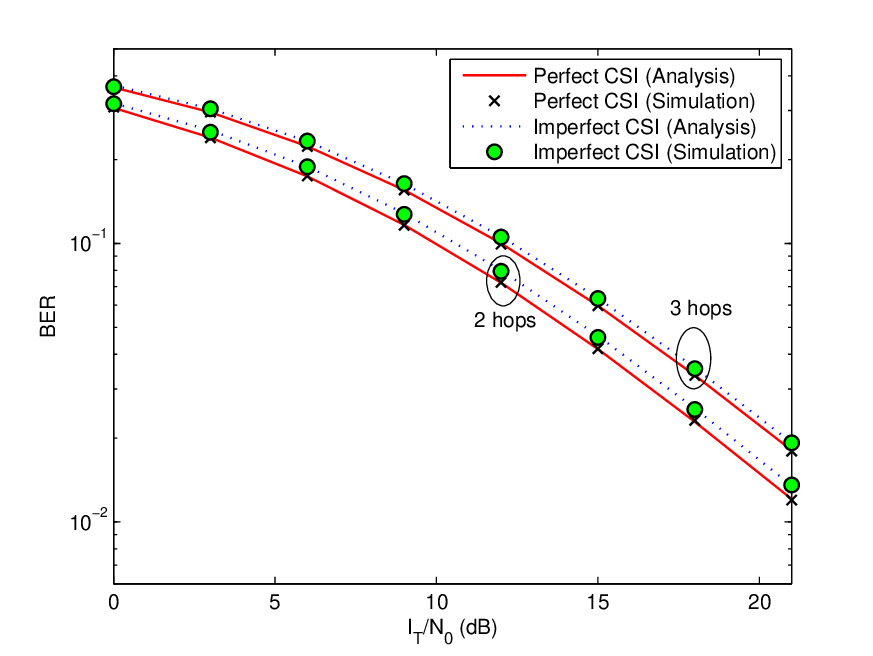}
\caption{BER versus $I_T/N_0$ (4-QAM).} \label{Figure14QAM}
\end{figure}

\begin{figure}
\centering
\includegraphics[height=9.5cm, width=9.5cm]{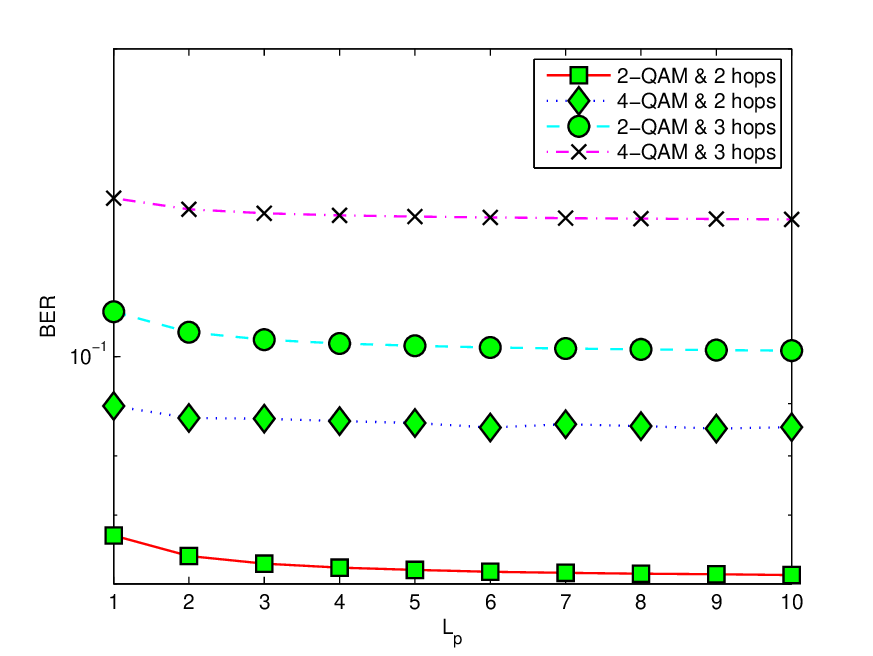}
\caption{BER versus $L_p$ ($I_T/N_0 = 10, \rho=1$dB).} \label{Fig:2}
\end{figure}

For illustration purpose, we arbitrarily select user coordinates as
shown in Fig.~\ref{Fig:Sim}: P at $(0.7, 0.5)$, SS 0 at $(0, 0)$, SR
1 at $(0.6, 0.2)$, SR 2 at $(0.8, 0.3)$, SD 3 at $(1, 0)$. SS 0, SD
3, and P are always fixed and thus, for 2-hop case only SR 1 is
considered. Also, the number on the line is the distance between two
corresponding terminals. The network topology in Fig.~\ref{Fig:Sim}
is applied to all following results.

We consider the path-loss exponent of $\alpha = 3$ and the CSI error
variance of $\sigma _{tr} = 1/\left( {{L_p}{B_{t,training}}{\eta
_{tr}}/{N_0} + 1} \right)$, \cite{Amin}. The value of
$B_{t,training}$ is selected such that the average received power at
P does not exceed $I_T$ (i.e., ${B_{t,training}}{\eta _{tr}}\leq
I_T$)\footnote{The study of channel estimators is outside the scope
of this paper. Therefore, the selection of $B_{t,training}$ in this
paper is just an example to demonstrate the effect of CSI
imperfection on the BER of underlay relay cognitive networks.}. As a
result, for illustration purposes we select $B_{t,training}=I_T/\eta
_{tP}$.

Figs.~\ref{Figure12QAM} and ~\ref{Figure14QAM} compare simulated
and numerical results for two typical modulation levels, namely,  $2$-QAM for
odd $q$ and $4$-QAM for even $q$, $N=\{2,3\}$, and different degrees
of CSI availability  - perfect CSI and imperfect CSI with $L_p = 1$.
It is seen that analytical results are well matched with simulated
ones, validating the derived expression. Additionally, the BER
performance is improved with respect to the increase in $I_T$. This
is obvious since $I_T$ imposes a constraint on the transmit power
and the higher $I_T$, the higher the transmit power, eventually
enhancing communication reliability. Moreover, the BER performance
is deteriorated with the lack of CSI.

Fig.~\ref{Fig:2} investigates the impact of the quality of the
channel estimator on the BER. The quality of the channel estimator
can be enhanced by increasing the number of pilot symbols $L_p$ at
the cost of the bandwidth loss due to increased overhead. The
results are reasonable since the BER performance is improved with
the increased $L_p$. Furthermore, for the selected channel estimator
model, the performance is saturated at $L_p = 4$.

Given the specific network topology in Fig.~\ref{Fig:Sim}, the
results in Figs.~\ref{Figure12QAM}, ~\ref{Figure14QAM},
and~\ref{Fig:2} illustrate that 3-hop communication is worst than
2-hop communication for any set $\{L_p, \alpha, I_T, M\}$. This
means that in underlay DF multi-hop cognitive networks the advantage
of the 3-hop communication over 2-hop communication in terms of the
path loss reduction, e.g., the distance from the last relay to the
destination in the 3-hop case (SR 2) is smaller than that in the
2-hop case (SR 1), can not sometimes turn into the performance
improvement. This is because the last relay in the 3-hop case is
closer to the primary user than in the 2-hop case, causing higher
interference. Thus, the last relay in the 3-hop case should utilize
lower transmit power than in the 2-hop case for reducing the
interference level to the primary user, leading to higher
performance degradation. These results recommend that the relay
selection in underlay DF multi-hop cognitive networks is crucial in
enhancing the network performance. A good relay not only provides
reliable communication to the destination but also causes less
interference to the primary user. The problem of the relay selection
will be considered in a future work.

\section{Conclusion}
\label{Conclusion}

This paper investigated analytically the BER performance of underlay DF multi-hop
cognitive networks over Rayleigh fading channel in consideration of
imperfect CSI. The derived expression was shown to have a convenient algebraic form which allows straightforward to timely evaluation of the corresponding performance. The proposed analytical results were supported and validated with results from computer simulations while
various results demonstrated that the imperfect CSI
affects significantly the BER of underlay DF multi-hop cognitive
networks. In addition, it was shown that the BER performance is dependent upon both the
number of hops and the network topology.


\end{document}